\documentclass[11pt]{article}
\usepackage{natbib}
\usepackage{graphicx}
\usepackage{amsmath}

 \usepackage[right,mathlines]{lineno} 
 \nolinenumbers

\date{}

\title{Determination of the Joint Confidence Region
of Optimal Operating Conditions in Robust Design
by Bootstrap Technique}
\begin{document}
\maketitle
\begin{abstract}
\begin{small}
Robust design has been
widely recognized as a leading method in reducing variability and
improving quality.
Most of the engineering statistics literature mainly 
focuses on finding \textit{point estimates} of the optimum 
operating conditions for robust design.
Various procedures
for calculating point estimates of the optimum operating conditions
are considered. 
Although this point estimation procedure is
important for continuous quality improvement, the immediate
question is ``how accurate are these optimum operating conditions?"
The answer for this is to consider interval estimation for a single variable
or joint confidence regions for multiple variables. 

In this paper, with the help of the bootstrap technique,
we develop procedures for obtaining joint \textit{confidence regions}
for the optimum operating conditions.
Two different procedures using Bonferroni 
and multivariate normal approximation are introduced.
The proposed methods are illustrated and substantiated using a numerical
example.

\end{small}
\end{abstract}

\section{Introduction}
Robust design has been
widely recognized as a leading method in reducing variability in the quality
characteristic and improving quality.
It is also recognized that quality improvement activities are most
efficient and cost-effective when implemented during the design stage.
Because of their practicability, 
robust design techniques have found increased applications
in many manufacturing industries.
Many industries are of great interest in the potential for applying
robust design principles and are seeking a role in the information
revolution with robust design at its core.

The primary goal of robust design is to minimize variation in the quality
characteristic of interest while keeping a process mean at the
customer-identified target value. 
In order to achieve this goal, 
\cite{Taguchi:1986} introduced a systematic method for applying
experimental design, which has become known as robust design. Even though
the \textit{ad hoc} robust design methods suggested 
by Taguchi remain controversial
due to various mathematical imperfections, there is no serious disagreement
among engineering researchers and practitioners about his basic philosophy. 
The controversy surrounding Taguchi's assumptions, his experimental design, 
and his experimental  analysis has been well addressed by 
\cite{Leon/Shoemaker/Kackar:1987}, 
\cite{Box:1988}, \cite{Box/Bisgaard/Fung:1988}, 
\cite{Nair:1992}, and \cite{Tsui:1992}. 
Consequently, researchers have closely examined alternatives 
using well-established
statistical tools from traditional theories of experimental designs. 
In an early attempt of such research, \cite{Vining/Myers:1990} introduced
the dual response approach based on a response surface methodology
(RSM) as a superior alternative for modeling process relationships
by separately estimating the response functions of the process mean
and variance. 
The mean target value and variance of the target
value are assumed to be polynomial functions of the various possible
design points. 
Multiple experiments are conducted at the various
design points in order to obtain estimates of the mean and variance
of the target value at the various design points. Then, given these
estimates,  the coefficients for the response surface functions are
estimated using standard regression techniques and the functions are
assumed to hold continuously and therefore for points between the design
points considered during the experiments. 
\cite{Vining/Myers:1990} then obtain
the optimal design point (usually somewhere between those used during
the experiment) by minimizing an objective function that penalizes
mean bias and variance.
Thus, it achieves the primary goal of robust design by
minimizing the process variance while adjusting the process mean at the
target. 

However, \cite{Lin/Tu:1995}
pointed out that the robust design solutions obtained
from the dual response model may not necessarily be optimal since this
model forces the process mean to be located at the target value and  proposed
the mean-squared-error model, relaxing the zero-bias assumption. While
allowing some process bias, the resulting process variance is less than
or at most equal to the variance obtained from the model proposed by
\cite{Vining/Myers:1990}; 
hence, the mean-squared-error model may provide better (or at
least equal) robust design solutions 
if the zero-bias assumption is not required. 
The robust design approach to determining optimum values 
has been further studied by several
researchers, including \cite{Borror/Montgomery:2000}, \cite{Scibilia/etc:2003},
\cite{Govindaluri/Shin/Cho:2004}, 
\cite{Lee/etc:2005}, \cite{Kim/etc:2005}, 
\cite{Lee/Park:2006a}, \cite{Lee/etc:2007}, and \cite{Lee/Park/Cho:2007a}.

As afore-mentioned, 
the majority of the research on robust design modelling have focused
on methodological development of models for optimizing operating conditions.
Although this is
critically important for continuous quality improvement, the immediate
question is ``how accurate are these optimum operating conditions?"
To the best of our knowledge,
the answer for this question has not been properly addressed 
in industrial applications of dual response surface methods.
We can consider interval estimation for a single variable
or joint confidence regions for multiple variables as an answer for this question.

With the help of the bootstrap technique,
we develop procedures for obtaining joint confidence regions
for the optimum operating conditions.
The single response surface approach uses the method
of least squares to obtain the adequate single response functions,
while the squared-loss approach uses two surfaces for process mean and variance.
For the single response surface approach,
the variance-covariance matrix of the regression
coefficients can be easily obtained,
so the derivation of the joint confidence region for optimum conditions
is straightforward.
For more details,
see \cite{Myers/Montgomery:2002}.
With two response functions, however, it is quite difficult to
obtain the variance-covariance matrix of the regression
coefficients for both the mean and variance responses,
particularly when the sample size is not large.
This difficulty can be overcome by using bootstrap techniques.
In an era of powerful computers, computer-intensive methods
such as the bootstrap technique promise to be one of the mainstays of
applied response surface methodology and engineering statistics
in the years ahead.

This paper is organized as follows.
The basics of robust design are  introduced in Section~\ref{SEC:RD}.
The bootstrap approach to determining joint confidence region for 
the optimum conditions is described in Section~\ref{SEC:BOOT},
followed by 
a case study in Section~\ref{SEC:Case}
and the paper ends with concluding remarks in Section~\ref{SEC:Concluding}.

\section{Robust design based on response surface}\label{SEC:RD} 
We consider a system with a response $Y$.
This response depends on the levels
of $k$ control factors $\mathbf{x}=(x_1,x_2,\ldots,x_k)$.
The following assumptions are generally made:
\begin{itemize} 
\item[(i)] The response $Y$ depends on $\mathbf{x}$. 
           Thus, $Y$ can be viewed as a function of $\mathbf{x}$, that is,
          $Y=F(x_1,x_2,\ldots,x_k)$.
 But, the functional structure $F(\cdot)$ is either unknown or very complicated.
\item[(ii)] The levels of the control factor $x_i$ for $i=1,2,\ldots,k$ 
  are continuously quantitative.
\item[(iii)] The levels of the control factor $x_i$ for $i=1,2,\ldots,k$ can be controlled by
      the experimenter.
\end{itemize}

Following \cite{Vining/Myers:1990},
the mean and variance response functions (surfaces) can be written as
 \begin{linenomath}
\begin{align*}
{M}(\mathbf{x})
&= {\beta}_0 + \sum_{i=1}^{k}{\beta}_i x_i + \sum_{i=1}^{k}{\beta}_{ii} x_i^2
             + \sum_{i<j}^{k} {\beta}_{ij} x_i x_j + \epsilon_m \\
\intertext{and} 
{V}(\mathbf{x})
&= {\gamma}_0 + \sum_{i=1}^{k}{\gamma}_i x_i +  \sum_{i=1}^{k}{\gamma}_{ii} x_i^2
              + \sum_{i<j}^{k}{\gamma}_{ij} x_i x_j + \epsilon_v,
\end{align*}
 \end{linenomath}
respectively, where $\mathbf{x}=(x_1, x_2, \ldots, x_k)$.
To find the fitted response surfaces given above, we must regress the mean response
${M}(\mathbf{x})$ and the variance response ${V}(\mathbf{x})$
on the control factors, $x_1, x_2, \ldots, x_k$.
Hence, we must estimate the mean and variance responses at each design point.
The most popular estimation method is to find the maximum likelihood estimates,
 assuming that
the error variables $\epsilon_m$ and $\epsilon_v$ are normally distributed.
Suppose that $n$ replicates are taken at the $i$th design point.
Let $Y_{ij}$ represent the $j$th response at the $i$th design point
where $i=1,2,\ldots,k$ and $j=1,2,\ldots,n$.
The most popular estimators of the location and scale parameters
are mean and variance, respectively.
The maximum likelihood estimates of the mean and variance
at the $i$th design point are the sample mean and variance as shown below.
\begin{linenomath}
$$
\overline{Y}_i = \frac{1}{n}\sum_{j=1}^{n} Y_{ij}
\mathrm{~~and~~}
S^2_i = \frac{1}{n-1} \sum_{j=1}^{n} (Y_{ij}-\overline{Y}_i)^2.
$$
\end{linenomath}

Let $\hat{M}(\mathbf{x})$ and $\hat{V}(\mathbf{x})$ represent
the fitted response functions for the mean and variance of the response $Y$,
respectively.
Assuming a second-order polynomial model for the response functions, we get
\begin{linenomath}
\begin{equation}  \label{EQ:m}
\hat{M}(\mathbf{x}) = \hat{\beta}_0 + \sum_{i=1}^{k}\hat{\beta}_i x_i
            + \sum_{i=1}^{k}\sum_{j=i}^{k}\hat{\beta}_{ij} x_i x_j
\end{equation}
\end{linenomath}
and
\begin{linenomath}
\begin{equation}  \label{EQ:v}
\hat{V}(\mathbf{x}) = \hat{\gamma}_0 + \sum_{i=1}^{k}\hat{\gamma}_i x_i
            + \sum_{i=1}^{k}\sum_{j=i}^{k}\hat{\gamma}_{ij} x_i x_j.
\end{equation}
\end{linenomath}
We use the sample mean and variance of $Y$ 
to estimate the process mean $\hat{M}(\mathbf{x})$ 
and variance $\hat{V}(\mathbf{x})$, respectively.

The main goal of robust design is to obtain   
the optimum operating conditions of control factors, $x_1, x_2, \ldots, x_k$,
and this can be achieved 
by employing the following squared-loss optimization model:
\begin{linenomath}
$$
\mathrm{minimize~} \big\{\hat{M}(\mathbf{x})-T_0\big\}^2 + \hat{V}(\mathbf{x})
$$
subject to  
$$
x_j \in [L_j,U_j] \mathrm{~~for~} j=1, 2, \ldots,k,
$$
\end{linenomath}
where $T_0$ is the customer-identified target value for the quality
characteristic of interest, and the constraint specifies the feasible joint region
of operating covariates given by $\mathbf{x}=(x_1, x_2, \ldots, x_k)$. 
When factorial designs with $k$ levels are used, the constraint becomes
$x_j \in  [L_j,U_j] \mathrm{~for~} j=1, 2, \ldots,k$.
The control factors ($x_j$) solving the optimization problem above 
are the optimal design point estimates.

When considering the process variance 
on the left hand side of the regression model in (\ref{EQ:v}), 
one often uses the log-transformed values of the sample
variances, {\em i.e.},  $\log(S_i^2)$, 
since a linear model for the variance process does not guarantee that
the predicted values are always positive;
see \cite{Myers/Montgomery:2002}.
Using the log-transform, we can avoid the problem of the negative estimates in the variance 
process. 
After estimating the logarithm of the process variance, 
${V}_{\log}(\mathbf{x})$, we can 
obtain the optimal operating conditions  
by minimizing 
\begin{linenomath}
$$
\{\hat{M}(\mathbf{x})-T_0\}^2 + \exp\big(\hat{V}_{\log}(\mathbf{x})\big)
$$
\end{linenomath}
subject to 
$$
x_j \in [L_j,U_j] \textrm{~~for~~} j=1, 2, \ldots,k .
$$

It is noteworthy that 
the following dual-response optimization model proposed by 
\cite{Vining/Myers:1990}    
can also be used for optimization purposes:
\begin{linenomath}
$$
\mathrm{minimize~} \hat{V}(\mathbf{x})
$$
\end{linenomath}
subject to  
$$\hat{M}(\mathbf{x}) = T_0 \textrm{~~and~~}
x_j \in [L_j,U_j] \textrm{~~for~~} 
j=1, 2, \ldots,k .
$$
However, the dual-response model strictly imposes a zero-bias condition
while the squared-loss model allows some bias 
({\em i.e.}, absolute value of the difference of $\hat{M}(x)$ and $T_0$).
This squared-loss model often results in less variability,
and will be the focus of this paper. 
For detailed information regarding the squared-loss model,
readers may refer to \cite{Lin/Tu:1995}. 

\section{Joint Confidence Region using Bootstrapped Samples}\label{SEC:BOOT}
The bootstrap technique was first developed by 
\cite{Efron:1979} 
and this method has become 
one of the most popular computer-intensive statistical methods.
The technique is simple yet powerful.
The key idea is to retake samples from the original data in order to create 
re-sampled data sets from which the variability of the quantities of interest
can be assessed without long and error-prone analytical calculations.

As afore-mentioned, unlike the single response surface method, 
it is quite difficult to
obtain the theoretical variance-covariance matrix of the regression 
coefficients for both the mean and variance responses when
two response functions are considered.
The alternative is to calculate the statistic of interest from simulated
data sets using the Bootstrap re-sampling technique.
We denote such a simulated data set as $Y_{ij}^*$.
It is standard to let the superscript notation (${}^*$) 
denote a bootstrapped or re-sampled value.
The statistic of interest (optimum operating conditions) is calculated 
with a simulated data set.
By simulating $B$ times, 
we obtain $B$ simulated optimum operating conditions.
Using these conditions, we can obtain the joint confidence region of 
the optimum operating conditions of control factors.
The bootstrap algorithm is as follows:

For $b=1,2,\ldots, B$:
\begin{enumerate}
\item[1] Draw $Y_{i1}^*,Y_{i2}^*,\ldots,Y_{in}^*$ with replacement from 
           $Y_{i1},Y_{i2},\ldots,Y_{in}$ for $i=1,2,\ldots,k$.
\item[2] Find the optimum operating conditions using the above data set.
         We denote this as 
         $(\hat{x}_{1,b}^{\mathrm{oc*}},\hat{x}_{2,b}^{\mathrm{oc*}})$.
\item[3] Repeat the above two steps for $b=1,2,\ldots, B$.
\end{enumerate}
Then we obtain $B$ simulated optimum operating conditions 
$(\hat{x}_{1,b}^{\mathrm{oc*}},\hat{x}_{2,b}^{\mathrm{oc*}})$ 
 for $b=1,2,\ldots, B$.
There are several ways of finding the joint confidence regions 
using bootstrapped samples.
For more details, the reader is referred to
\cite{Hall:1992} 
and 
\cite{Davison/Hinkley:1997}.
Here we briefly introduce two methods: 
\begin{enumerate}
\item[(i)]  The construction of a rectangular region using a Bonferroni argument.
\item[(ii)] The construction of an elliptical contour based on an approximation 
to the multivariate normal distribution. 
\end{enumerate}
In the following section, we describe these two methods
using a numerical example with multi-filament microfiber tows.

\section{A Case Study}\label{SEC:Case}
A company produces multi-filament microfiber tows.
We conduct an experiment to determine the quality effect of control covariates.
For such products, the key control factors are polymer temperature ($x_{1,i}$) 
and polymer feeding speed ($x_{2,i}$). 
The diameter ($Y$) of the microfiber is the most important quality issue
and its nominal target value is $T_0=50$ microns.
The $3\times3$ factorial design taken at each design point 
for $i = 1,2,\ldots,9$ and $j=1,2,\ldots,10$ is shown in Table~\ref{TBL:Example1}.
Here, the variables $x_{1,i}$ and $x_{2,i}$ are centered and re-scaled from the natural variables
so that $x_{1,i}$ and $x_{2,i}$ are in $[-1,1]$.
We then obtain the estimate of the mean response, $\hat{M}(\mathbf{x})$, 
and the estimate of the log-transformed variance response, 
$\hat{V}_{\log}(\mathbf{x})$, 
as follows. 
\begin{linenomath}
\begin{align*}
\hat{M}(\mathbf{x})& = 
 51.741 + 7.750x_1 + 8.053x_2 + 20.262x_1^2 + 19.939x_2^2   - 0.038x_1x_2. \\ 
\hat{V}_{\log}(\mathbf{x})& =
 0.841  -0.015x_1   -0.068x_2 + 0.620x_1^2  + 0.421x_2^2  -0.339 x_1x_2. 
\end{align*}
\end{linenomath}
By minimizing 
$$
\{\hat{M}(\mathbf{x})-50\}^2 + \exp\big(\hat{V}_{\log}(\mathbf{x})\big)
$$
subject to  
$$
|x_1| \le 1  \textrm{~~and~~} |x_2| \le 1,
$$
the optimum operating conditions are obtained as 
$$
\hat{\mathbf{x}}^{\mathrm{oc}}
=(\hat{x}_1^{\mathrm{oc}},\hat{x}_2^{\mathrm{oc}})=( -0.168, -0.179). 
$$

\setlength{\doublerulesep}{\arrayrulewidth}
\begin{table}[t!]
\renewcommand{\arraystretch}{0.70}
\begin{center}
\caption{\label{TBL:Example1} Data for case study example.}
\medskip
\begin{tabular}{ccrrcrrrrrc@{~}rc@{~}r} 
\hline\hline\hline \\[-2.5ex]
$i$ && $x_{1,i}$ & $x_{2,i}$  && \multicolumn{5}{c}{$Y_{ij}$} &&
$\overline{Y}_i$ && $S^2_i$ \\[0.5ex]
\cline{1-1} \cline{3-4} \cline{6-10} \cline{12-12} \cline{14-14} 
1&&$-1$&$-1$&&  73.94& 76.09& 73.39& 79.82& 76.47& \\
 &&    &    &&  73.43& 76.89& 77.55& 77.12& 74.79&& 75.949&& 4.263\\
2&&$ 0$&$-1$&&  67.30& 64.55& 62.08& 58.18& 66.36& \\
 &&    &    &&  63.49& 63.56& 65.91& 65.61& 65.05&& 64.209&& 6.853\\
3&&$ 1$&$-1$&&  94.03& 93.67& 91.80& 86.34& 93.24& \\
 &&    &    &&  91.45& 91.19& 87.71& 90.33& 92.71&& 91.247&& 6.390\\
4&&$-1$&$ 0$&&  66.93& 63.35& 64.55& 63.47& 60.23& \\
 &&    &    &&  62.58& 62.63& 63.45& 66.29& 65.47&& 63.895&& 3.922\\
5&&$ 0$&$ 0$&&  51.23& 51.03& 53.16& 52.84& 50.06& \\
 &&    &    &&  50.02& 52.42& 53.32& 51.35& 53.57&& 51.900&& 1.775\\
6&&$ 1$&$ 0$&&  80.58& 78.10& 80.44& 76.83& 83.11& \\
 &&    &    &&  84.45& 78.70& 77.04& 81.00& 79.27&& 79.952&& 6.181\\
7&&$-1$&$ 1$&&  97.95& 91.50& 93.42& 91.67& 89.63& \\
 &&    &    &&  92.10& 86.82& 95.48& 92.01& 97.35&& 92.793&&11.631\\
8&&$ 0$&$ 1$&&  80.76& 77.86& 81.10& 77.31& 76.53& \\
 &&    &    &&  80.31& 78.51& 79.60& 79.78& 78.16&& 78.992&& 2.377\\
9&&$ 1$&$ 1$&& 106.10&107.24&110.72&103.57&109.17& \\
 &&    &    && 108.48&110.41&106.80&108.58&108.31&&107.938&& 4.495\\
\hline\hline\hline  \\
\end{tabular}
\end{center}
\end{table}

Two methods,
the construction of a rectangular confidence region using a Bonferroni argument 
and the construction of an elliptical confidence region 
based on a multivariate normal distribution,
are used to find the joint confidence region for the optimum operating conditions.
For each bootstrap data set $Y^*_{ij}$ where $i=1,2,\ldots,9$ and 
$j=1,2,\ldots,10$,
we obtain the following simulated estimates of the optimum conditions
as shown in Table~\ref{TBL:Boot1}.

\bigskip 
\setlength{\doublerulesep}{\arrayrulewidth}
\begin{table}[t!]
\renewcommand{\arraystretch}{1.00}
\begin{center}
\caption{\label{TBL:Boot1} Simulated estimates of the optimum conditions
from bootstrap data sets.}
\medskip
\begin{tabular}{@{}crrrrrrrrr} 
\hline\hline\hline \\[-2.5ex]
$b$ & 1  & 2  & 3 & 4 & 5 & 6 & $\cdots$ & 999  \\
\hline 
$\hat{x}^{\mathrm{oc}*}_{1,b}$               
&$-0.268$ &$-0.254$& $-0.154$& $-0.079$& $-0.207$&$-0.202$& $\cdots$ & $-0.169$ \\
$\hat{x}^{\mathrm{oc}*}_{2,b}$               
&$-0.141$ &$-0.089$& $-0.207$& $-0.146$& $-0.207$&$-0.197$& $\cdots$ & $-0.196$ \\
$\hat{M}(\hat{\mathbf{x}}^{\mathrm{oc}*}_b)$ 
&$50.024$ &$50.060$& $50.255$& $50.036$& $50.692$&$50.399$& $\cdots$ & $50.503$ \\
\hline\hline\hline   \\
\end{tabular}
\end{center}
\end{table}

A Bonferroni argument can be used to find  the rectangular confidence region.
Suppose that $\theta$ is $d$-dimensional and
the joint confidence region 
$C^{\alpha}=(C^{\alpha_1}_1, C^{\alpha_2}_2, \ldots, C^{\alpha_d}_d)$
is rectangular,
with the interval $C^{\alpha_i}_i=(\theta^L_i,\theta^U_i)$ for the $i$th component,
where $i=1,2,\ldots,d$.
Then we have
\begin{linenomath}
$$
P(\theta \not\in C^{\alpha})
= P\big({\bigcup_{i=1}^{d}}\{ \theta_i \not\in C^{\alpha_i}_i\} \big)
\le \sum_{i=1}^{d} P(\theta_i \not\in C^{\alpha_i}_i) 
=   \sum_{i=1}^{d} \alpha_i.
$$
\end{linenomath}
If we take $\alpha_i=\alpha/d$, 
then the region $C^{\alpha}$ covers  at least $100(1-\alpha)$\%.
In our particular example, the region is two-dimensional, so we have $d=2$. 
If we want to find the 90\% joint confidence region, we can use
$\alpha_1=0.05$ and $\alpha_2=0.05$.
To obtain the joint confidence, we need to 
find the confidence interval $C^{\alpha_i}_i$ for $i=1,2$.
That is, we find the 95\% confidence intervals of 
${x}^{\mathrm{oc}}_{1,b}$ and ${x}^{\mathrm{oc}}_{2,b}$, separately. 
First, we will find the interval of ${x}^{\mathrm{oc}}_{1,b}$.
For convenience, we denote ${x}^{\mathrm{oc}}_{1,b}$ as $\theta$ and 
the estimator of $\hat{x}^{\mathrm{oc}}_{1,b}$ as $T$.
The realization of the estimator $T$ is denoted by $t$,
and its bootstrap simulated values are denoted by  $t^*$.
To find the confidence interval,
we need to estimate quantiles
for $T-\theta$ and these are approximated using 
the bootstrap quantile of $t^*-t$.
The $p$th quantile of $T-\theta$ is estimated by
the $(B+1)p$th ordered value of $t^*-t$, that is $t^*_{[(B+1)p]}-t$. 
Then, as described in 
\cite{Davison/Hinkley:1997}, 
an equi-tailed
$100(1-\alpha_1)$\% confidence interval will have the following endpoints:
\begin{linenomath}
$$
t-(t^*_{[(B+1)(1-\alpha_1/2)]}-t) \mathrm{~~and~~} 
t-(t^*_{[(B+1)\alpha_1/2]}-t).
$$
\end{linenomath}
In this particular bootstrap simulation, we used $B=999$.
Thus, we have
\begin{linenomath}
$$
t^*_{[(B+1)(1-\alpha_1/2)]}=t^*_{[975]} \mathrm{~~and~~}
t^*_{[(B+1)(\alpha_1/2)]}=t^*_{[25]}.
$$
\end{linenomath}

\begin{figure}[tb]
\renewcommand{\arraystretch}{0.50}
\centering\includegraphics{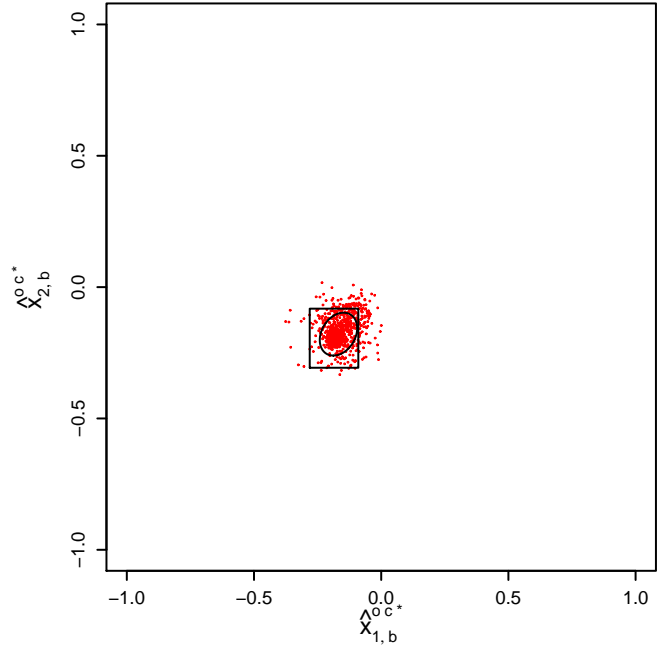}  
\caption{Scatter plot of bootstrap results of the optimum operation
conditions.} \label{FIG:plota}
\end{figure}

Similarly, we can find the confidence interval for ${x}^{\mathrm{oc}}_{2,b}$.
Although the Bonferroni argument has a long history in statistics, 
it is well known that the Bonferroni joint confidence region
is wider than what they should be at a given confidence level and it is therefore conservative.
Also, with plausible likelihood function contours, 
a circular or elliptic shape could be more appropriate than
a rectangular shape.
One possible simple remedy for these deficiencies of the Bonferroni method is
to use the classical normal-distribution approximation method.
In our example, the region is two-dimensional, so a joint confidence region
is an ellipse.
If the $d$-dimensional estimator $\mathbf{T}$ of $\boldsymbol{\theta}$ 
is approximately multivariate normal, 
then it is well known that the quadratic quantity $Q$ 
shown below has an approximate  $\chi^2_d$ distribution
\begin{linenomath}
$$
Q(\boldsymbol{\theta})
 = (\mathbf{T}-\boldsymbol{\theta})' \hat{\boldsymbol{\Sigma}}^{-1}
   (\mathbf{T}-\boldsymbol{\theta}),
$$
\end{linenomath}
where $\boldsymbol{\Sigma}$ is the estimated variance-covariance matrix
of $\mathbf{T}$.

\begin{figure}[h!]
\renewcommand{\arraystretch}{0.50}
\centering\includegraphics{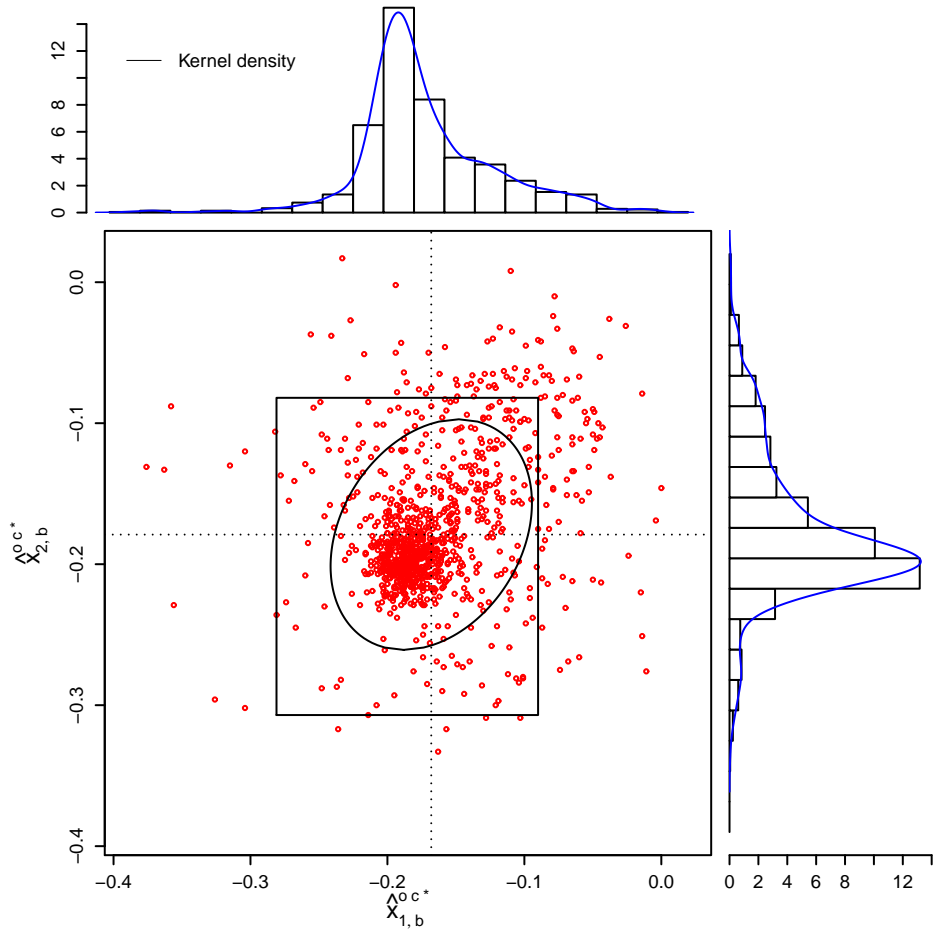}  
\caption{ Blowup plot of Figure~\ref{FIG:plota} with marginal 
histograms and kernel densities.}  
\label{FIG:plotb}  

\bigskip 
\bigskip 
\end{figure}

If $Q$ has exact chi-square quantiles $\chi_d^{2}(p)$, 
then a $100(1-\alpha)$\% joint confidence region for $\boldsymbol{\theta}$ is 
$$
\{\boldsymbol{\theta} : Q(\boldsymbol{\theta}) < \chi^2_{d}(1-\alpha) \}.
$$
However, such chi-square quantiles are often unreliable, so it makes sense
to use the bootstrap approximation of the distribution of $Q$ to find
quantiles of $Q$. 
The bootstrap analogue of $Q$ is
\begin{linenomath}
$$
Q^*(\mathbf{t})
 = (\mathbf{T}^*-\mathbf{t})' {\boldsymbol{\hat{\Sigma}}^*}^{-1}
   (\mathbf{T}^*-\mathbf{t}),
$$
\end{linenomath}
which is calculated for each of the $B$ bootstrap samples and its calculation
is denoted by $q^*$.
If we denote the ordered bootstrap values by 
$q^*_{[1]}, q^*_{[2]}, \ldots, q^*_{[B]}$, 
then $100(1-\alpha)$\% bootstrap  joint 
confidence region is given by 
\begin{linenomath}
$$
\{ \boldsymbol{\theta}: 
(\mathbf{t}-\boldsymbol{\theta})' {\boldsymbol{\hat{\Sigma}}}^{-1}
(\mathbf{t}-\boldsymbol{\theta}) < q^*_{[(B+1)(1-\alpha)]}
\}.
$$
\end{linenomath}
In the calculation of $Q^*$, we need to find
the variance estimator ${\boldsymbol{\hat{\Sigma}}^*}$. 
One general way to obtain the value for ${\boldsymbol{\hat{\Sigma}}^*}$
is to calculate
\begin{linenomath}
$$
{\boldsymbol{\hat{\Sigma}}^*}
=\frac{1}{I-1} \sum_{i=1}^I 
(\mathbf{t}_i^{**}-\mathbf{\bar{t}}^{**})
(\mathbf{t}_i^{**}-\mathbf{\bar{t}}^{**})',
$$
\end{linenomath}
where
$\mathbf{\bar{t}}^{**} = (1/I)\sum_{k=1}^{I}\mathbf{t}_k^{**}$
and 
 $\mathbf{t}_1^{**}, \mathbf{t}_2^{**}, \ldots, \mathbf{t}_I^{**}$ 
are calculated
by bootstrap {\it re-sampling} from the bootstrap sample for each of the $B$
simulated samples. 
Since $\mathbf{t}_i^{**}$ are obtained by using bootstrap re-sampling  
of the bootstrapped sample, the double superscript notation (${}^{**}$) is used.
Typically, $I$ is in the range between $50$ and $200$.
In this example, we set $I=100$.
For convenience, we denote 
$$
\boldsymbol{\theta}=({x}^{\mathrm{oc}}_{1},{x}^{\mathrm{oc}}_{2})' 
\textrm{~and~} 
\mathbf{T}=(\hat{x}^{\mathrm{oc}}_{1},\hat{x}^{\mathrm{oc}}_{2})' .
$$

Figures~\ref{FIG:plota} and \ref{FIG:plotb} show the
scatter plots of $B=999$ pairs of the optimum conditions based on the bootstrap samples. 
Figure~\ref{FIG:plotb} is the blowup plot of Figure~\ref{FIG:plota} with the
marginal histograms and kernel density estimates
 of $\hat{x}^{\mathrm{oc*}}_{1,b}$ and $\hat{x}^{\mathrm{oc*}}_{2,b}$.
The values of the optimum conditions from the original sample
is superimposed on the plot. The vertical dotted line
is    $\hat{x}^{\mathrm{oc}}_{1}=-0.168$ and
the horizontal dotted line is $\hat{x}^{\mathrm{oc}}_{2}=-0.179$.

The rectangular region is the 90\% joint  confidence region
based on Bonferroni simultaneous confidence intervals, 
while the 90\%  confidence elliptic region is based on the multivariate normal approximation. 
The bootstrap biases defined as 
$$
\frac{1}{B}\sum_{b=1}^B\hat{x}^{\mathrm{oc*}}_{i,b}-\hat{x}^{\mathrm{oc}}_{i}
$$
for $i=1, 2$ are  $0.00208$ and $0.00411$, respectively.
It is noteworthy that 
the rectangular confidence region deviates somewhat 
from the main body of the scatter plot downward and toward the left. 
This deviation comes from the bootstrap bias and skewness of the first and second components
of the optimum operating conditions.

\begin{figure}[t!]
\renewcommand{\arraystretch}{0.50}
 \centering\includegraphics{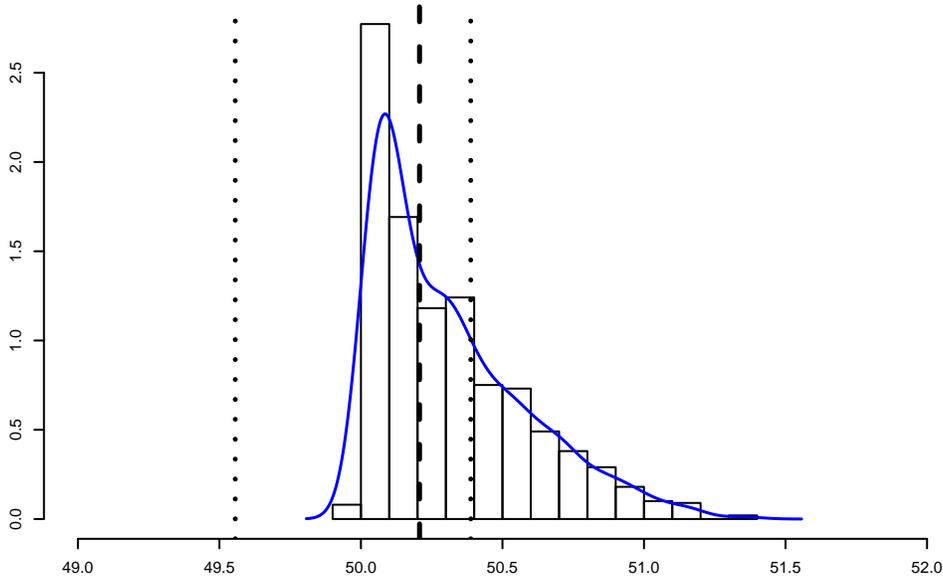}  
\caption{Histogram of the bootstrap mean response values.
\label{FIG:plotc}}
\end{figure}

We can also find the confidence interval of the \textit{mean response}
at the optimum conditions. 
Using this bootstrap technique, we can find the bootstrap estimates 
of the mean response at the optimum, that is, 
$\hat{M}(\hat{\mathbf{x}}^{\mathrm{oc}*}_b)$.
The mean response estimate at the optimum from the original sample is
$50.30$. 
Using the simulated bootstrap samples, we can find the 90\% bootstrap confidence interval,
$(49.36,50.59)$.
Figure~\ref{FIG:plotc} also shows the histogram of
bootstrap mean response values, where
the dotted vertical lines are the end points of the confidence interval
and the vertical dashed line is 
the mean response estimate at the optimum from the original sample.
The bias of the bootstrap estimate is 0.17.
Because of this bias, 
we observe that the confidence interval moves toward the left.

\section{Concluding remarks}\label{SEC:Concluding}
In this paper, we developed methods of constructing joint 
confidence regions for optimum operating conditions using the bootstrap
technique. Two different procedures were developed: A Bonferroni type
procedure that constructed a rectangular region and multivariate normal
approximation procedure that constructed an elliptical region. 
The proposed methods were illustrated and substantiated using a numerical
example involving multi-filament microfiber  tows.


\bibliographystyle{natbib}


\end{document}